\documentclass[prb,twocolumn]{revtex4-1}

\usepackage{amsmath}
\usepackage{amsfonts}
\usepackage{graphicx}

\usepackage[T1]{fontenc}  

\begin{document}

\title{Numerical matrix method for quantum periodic potentials}

\author{Felipe Le Vot}

\author{Juan J. Mel\'endez}
\email{melendez@unex.es}

\author{Santos B. Yuste}
\email{santos@unex.es}

\affiliation{Departamento de F\'{\i}sica and Instituto de Computaci\'on Cient\'{\i}fica Avanzada (ICCAEX), Universidad de Extremadura, 06006 Badajoz, Spain}

\date{\today}

\begin{abstract}
A numerical matrix methodology is applied to quantum problems with periodic potentials. The procedure consists essentially in replacing the true potential by an alternative one, restricted by an infinite square well, and in expressing the wave functions as finite superpositions of eigenfunctions of the infinite well. A matrix eigenvalue equation then yields the energy levels of the periodic potential within an acceptable accuracy. The methodology has been successfully used to deal with problems based on the well-known Kronig-Penney (KP) model.  Besides the original model, these problems are a dimerized KP solid, a KP solid containing a surface, and a KP solid under an external field. A short list of additional problems that can be solved with this procedure is presented.
\end{abstract}

\maketitle

\section{Introduction}
\label{secIntro}

The potentials appearing in the Schr\"{o}dinger equation as considered in undergraduate courses are traditionally expected to have two basic didactic properties: first, they should illustrate relevant physics (energy quantization, tunneling, features of the steady states, etc.)\ of real quantum systems; and second, the Schr\"{o}dinger equation should have analytical or semi-analytical solutions that, hopefully, can be worked out by the students. Unfortunately, the second condition severely reduces the set of suitable potentials, which hampers our ability to illustrate quantum phenomena.

It has been long recognized that this difficulty can be eased if numerical methods for solving the Schr\"{o}dinger equation are also employed.\cite{Kinderman1990} There are several numerical techniques that, while not excessively specialized, can deal with this equation. Most popular, perhaps, are finite-difference methods, which are easy to understand and employ in their simplest version.\cite{Kinderman1990,French1979,Chow1972,Johnston1992} These methods clearly show how the requirement of physically acceptable bound solutions leads to energy quantization.\cite{French1979} Recently, Marsiglio \textit{et al.}\ have described a quite different numerical procedure in which the Schr\"{o}dinger equation and its solutions are written in matrix and vector form, respectively, in the basis of the infinite square-well eigenfunctions. These authors have obtained approximate solutions for the Schr\"{o}dinger equation in one dimension (harmonic potential, finite square well,\cite{Marsiglio2009} and a set of periodic potentials\cite{Pavelich15}) and for the radial equation for three-dimensional potentials with spherical symmetry (Coulomb, Yukawa, and finite spherical well potentials).\cite{Jugdutt2013}

The matrix approach has some nice features. First, it helps students understand vector spaces and the matrix representation of quantum operators.  Typically, in undergraduate courses the matrix representation of quantum operators appears when discussing angular momentum (Pauli matrices, Clebsch-Gordan coefficients) and degenerate perturbation theory,\cite{Griffiths2004} but its connection to the previously well-studied Schr\"{o}dinger formalism is unclear. The approach of Refs.\ \onlinecite{Marsiglio2009,Pavelich15,Jugdutt2013} provides a simple way to relate the two formalisms.  It is especially fortunate that, in this approach, the solutions are expanded in the basis of the infinite square-well eigenfunctions because this is just a Fourier expansion, a topic many students are familiar with. Finally, this approach leads to a numerical method that is quite simple to use, accurate, valid for a large number of potentials (essentially, all potentials $V(x)$ for which the integral of $V(x)$ times the product of sinusoidal functions can be evaluated), and can be easily programmed, especially with modern software packages such as \textsc{Mathematica}, providing in many cases excellent results with quite modest computational cost. (As supplementary materials we include in Ref.~\onlinecite{SuplementaryMaterials} the \textsc{Mathematica} codes employed to solve the systems and examples considered in this paper.)

Another nice feature of the matrix approach is that it provides at once numerical estimates of the first $N$ energies and eigenfunctions of the Schr\"{o}dinger equation, where $N$ can be set at will. Compare this with the standard finite-difference approach, where energies and eigenfunctions are obtained one by one.\cite{Johnston1992} This property makes the matrix method especially suitable for the study of periodic potentials where many energies are involved in the formation of energy bands. Besides, in some cases these energies are so close that some of them, and their corresponding eigenfunctions, can be easily missed by standard finite-difference  methods. The matrix approach is free of this problem.

In this paper we exploit these characteristics of Marsiglio's matrix approach to the study of periodic potentials, with the Kronig-Penney (KP) model as archetype.\cite{Kronig1931} This model is commonly used in courses in solid-state physics to justify qualitatively the appearance of energy bands. There exist several procedures to solve the Schr\"{o}dinger equation with such a potential.\cite{Titus1973,Singh1982,Pavelich15} These procedures use the original approach of Kronig and Penney's paper, which starts from wave functions compatible with the Bloch theorem.
This means that they require some knowledge about solid-state physics, which most undergraduates have not reached when they study quantum physics. The method that we present here, on the contrary, does not require any background in solid-state physics; the energy bands arise naturally from the formalism. Thus, this method can ease the difficulties that students face when they extend their knowledge about quantum physics to crystalline solids.

In Sec.~\ref{secMatrixMet} we present the matrix formalism and point out its convenience for dealing with periodic potentials when a sufficiently large number of unit cells (periods) are considered. In Sec.~\ref{secPerPot} we apply this method to the study of the original KP model, a dimerized KP solid, a KP solid containing a surface, and a KP solid under an external field. The usefulness of this method to provide the time evolution of quantum systems is illustrated with an example involving the KP solid under an external field.

We note here that there are two especially useful complementary references to the present paper. One is Ref.~\onlinecite{Johnston1992}, in which the original KP model, the KP solid under an external field, and some other interesting variations of the KP model (doped lattices and amorphous lattices) are studied numerically using a finite-difference method.  The other is Ref.~\onlinecite{Pavelich15}, where Marsiglio's approach is used to obtain numerical solutions within a single unit cell, which are then extended by means of Bloch's theorem to periodic potentials.

\section{The matrix method}
\label{secMatrixMet}

Let $H_0$ be the Hamiltonian for an infinite square-well potential (box) of width $L$:
\begin{equation}
H_0=-\frac{\hbar^2}{2\mu}\,\frac{d^2}{dx^2}+V_\text{inf}(x),
\end{equation}
where
\begin{equation}
V_\text{inf}(x)=
\begin{cases}
0, & 0\le x\le L\\
\infty,&\text{otherwise}
\end{cases}
\end{equation}
and $\mu$ is the mass of the particle. The eigenfunctions of $H_0$ are
\begin{equation}
\label{1}
\varphi_p(x) =
\begin{cases}
\sqrt{\dfrac{2}{L}}\, \sin\left(\dfrac{p\pi x}{L}\right), & 0\le x\le L\\
0,&\text{otherwise}
\end{cases}
\end{equation}
and the corresponding eigenvalues are
\begin{equation}
\label{2}
E_p^\text{(0)}=
\frac{\hbar^2\pi^2 p^2}{2\mu L^2 },
\end{equation}
with $p=1,2,\ldots\,$.

Let us now denote by $\widetilde H$ the Hamiltonian
\begin{equation}
\widetilde H=-\frac{\hbar^2}{2\mu}\,\frac{d^2}{dx^2}+V(x)
\end{equation}
of the Schr\"{o}dinger equation
\begin{equation}
\label{wideH}
\widetilde H |\psi\rangle=E|\psi\rangle
\end{equation}
that we want to solve. Note that the potential $V(x)$ is not necessarily limited to the domain $0 \le x \le L$.

\begin{figure}[h!]
\centering
\includegraphics[width=8.5cm]{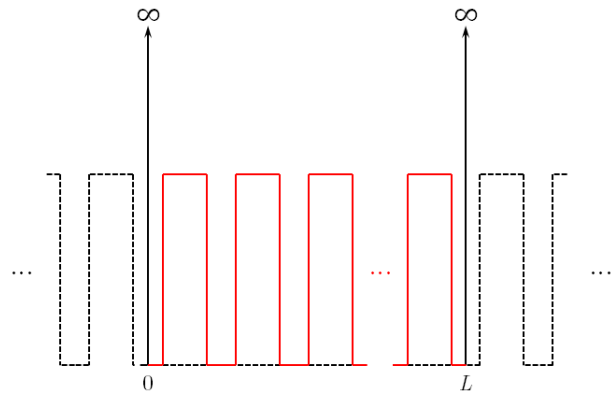}
\caption{The true potential $V(x)$ (dashed line) and the auxiliary potential $V_\text{inf}(x)+V(x)$ (solid line) for the Kronig-Penney model.
}
\label{fig1}
\end{figure}

The first approximation in the matrix method consists in replacing the solutions of Eq.~\eqref{wideH} with those of
\begin{equation}
\label{3}
H |\psi\rangle=E|\psi\rangle,
\end{equation}
where
 \begin{equation}
 H=H_0+V=-\frac{\hbar^2}{2\mu}\frac{d^2}{dx^2}+V_\text{inf}(x)+V(x).
 \end{equation}
Equation~\eqref{3} is just Eq.~\eqref{wideH} but with $V(x)$ replaced by $V_\text{inf}(x)+V(x)$.
This approximation is valid, for example, when the eigenfunctions $\psi(x)$ of $\widetilde H$ are negligible outside the box.\cite{Marsiglio2009} However, in this paper we will use this procedure to study periodic potentials, a case where the above condition does not hold. Specifically, we will replace the infinitely repeating potential $V(x)$ with a finite periodic potential $V_\text{inf}(x)+V(x)$ encompassing just a few periods inside the region $0 \le x \le L$ (see Fig.~\ref{fig1}). Although the wave functions of the (fully) periodic $\widetilde H$ are not at all negligible outside the box, our rationale here is that the effects of the box boundaries on some quantities (e.g., the allowed energies) will be negligible if the number of periods inside the box is large enough.  We will see that one can get an excellent qualitative and even quantitative description of true periodic systems by enclosing just a few periods within the box.

The next step is to note that the eigenfunctions $\{\varphi_p(x)\}$ of $H_0$ form a complete set of basis states for $H$, so we can express any solution $\psi(x)$ of the Schr\"{o}dinger equation \eqref{3} as the Fourier series
\begin{equation}
\label{4}
|\psi\rangle=\sum_{m=1}^\infty c_m |\varphi_m\rangle,
\end{equation}
where $\{c_m\}$ is a list of undetermined Fourier coefficients. Inserting Eq.~\eqref{4} into Eq.~\eqref{3} and
using the orthonormality of the $\{\varphi_p(x)\}$ functions, we obtain the eigenvalue equation in matrix form:
\begin{equation}
\label{matrixInfi}
\sum_{m=1}^\infty H_{nm}c_m =E c_n,
\end{equation}
where $H_{nm}=\langle\varphi_n |H| \varphi_m \rangle$ is the $nm$ matrix element of $H$ in the Fourier basis, that is,
\begin{equation}
\label{Hnmgral}
H_{nm}=\delta_{nm}\, E_n^{(0)}+
\frac{2}{L} \int_0^L\sin\Bigl(\frac{n\pi x}{L}\Bigr)V(x)\sin\Bigl(\frac{m\pi x}{L}\Bigr) \, dx ,
\end{equation}
with $n,m=1,2,\ldots$ and $\delta_{nm}$ being the Kronecker delta.

The matrix equation \eqref{matrixInfi} is fully equivalent to the Schr\"{o}dinger equation \eqref{3}, but is impractical because it is infinite in dimension.
Fortunately, not all the coefficients $\{c_m \}$ are required for an accurate representation of the wave function $\psi(x)$. As a second approximation, we therefore assume that it suffices to retain only a finite number $N$ of terms in Eq.~\eqref{4}, and to similarly truncate the sum in Eq.~\eqref{matrixInfi} at $m=N$.  The value of $N$ is chosen to obtain some predefined accuracy. In practice, for example, one starts with a trial value of $N$ and then increases this value until the effect of the increase on the energy eigenvalues is less than some desired threshold.

In summary, the numerical matrix method of Refs.\ \onlinecite{Marsiglio2009,Pavelich15,Jugdutt2013} consists of (1) embedding the potential within  an infinite square well; (2) expanding the wave function in the basis of eigenstates of the infinite square well, retaining only the first $N$ terms of the expansion, with $N$ chosen self-consistently; and (3) solving the eigensystem \eqref{matrixInfi}, truncated at dimension~$N$, to obtain the low-lying energies and the associated wave functions.

\section{Periodic potentials}
\label{secPerPot}

\subsection{Kronig-Penney model}
We study first the standard Kronig-Penney potential, which serves as a simple model of the periodic potential of a crystal.\cite{Kronig1931} We consider a one-dimensional crystal of lattice parameter $a$, where in each unit cell there is a centered barrier of width~$b$. The Kronig-Penney potential is then
\begin{equation}
\label{7}
V_{KP}(x) =
\begin{cases}
V_0, & \left|x-x_r\right| < b/2 \\
0,& \text{otherwise},
\end{cases}
\end{equation}
where $x_r=-a/2+ra$ is the position of the $r$th barrier, and $r$ is an integer. Analytical solutions for this potential are reported in the original Ref.~\onlinecite{Kronig1931} and elsewhere.\cite{Singh1982,Lippmann1997,Szmulowicz1997,Szmulowicz2008} All of these authors assume periodic boundary conditions, so that the Bloch theorem can be used.

Here we instead use the matrix approach described above, embedding the potential within an infinite square well of width $L$ so that the KP potential $V_{KP}$ is replaced by
\begin{equation}
\label{Vaux}
V_\text{inf}(x)+V_{KP}(x) =
\begin{cases}
\infty, & x\le 0 \text{ or } x\ge L\\
V_0, & \left|x-x_r\right| < b/2 \\
0,& \text{otherwise},
\end{cases}
\end{equation}
where $r=1,\ldots,n_b$ and $n_b=L/a$ is the number of barriers.
The matrix elements $H_{nm}=H^{KP}_{nm}$ given by Eq.~\eqref{Hnmgral} are then
\begin{align}
\label{HnnKP}
 H^{KP}_{nm}&= E_n^{(0)}\, \delta_{nm}+ V_0\,\sum_{r=1}^{n_b} h_{nm}(x_r,b)
\end{align}
with
\begin{equation}
\label{hnm}
h_{nm}(s,b)= \frac{2}{L} \int_{s-b/2}^{s+b/2}\,\sin\Bigl(\frac{n\pi x}{L}\Bigr) \sin\Bigl(\frac{m\pi x}{L}\Bigr) \, dx .
\end{equation}
This integral can be readily evaluated using trigonometric identities to obtain $h_{nm}(s,b)=F_{nm}(s+b/2)-F_{nm}(s-b/2)$,
with
\begin{align}
\label{Fnn}
F_{nn}(x)&= \frac{x}{L}-\frac{\sin( 2 \pi n x/L)}{2 \pi n}
\end{align}
and
\begin{align}
F_{nm}(x)&= \frac{\sin[(m-n)\pi x/L]}{\pi(m-n)}-\frac{\sin[(m+n)\pi x/L]}{\pi(m+n)}
\label{Fnm}
\end{align}
for $n\neq m$.
In what follows, we will use units such as $\hbar^2/2\mu=1$ and $a=1$,  which implies that energies are in units of $\hbar^2/ 2\mu a^2$.

\begin{figure}[h!]
\centering
    \includegraphics[width=8.5cm]{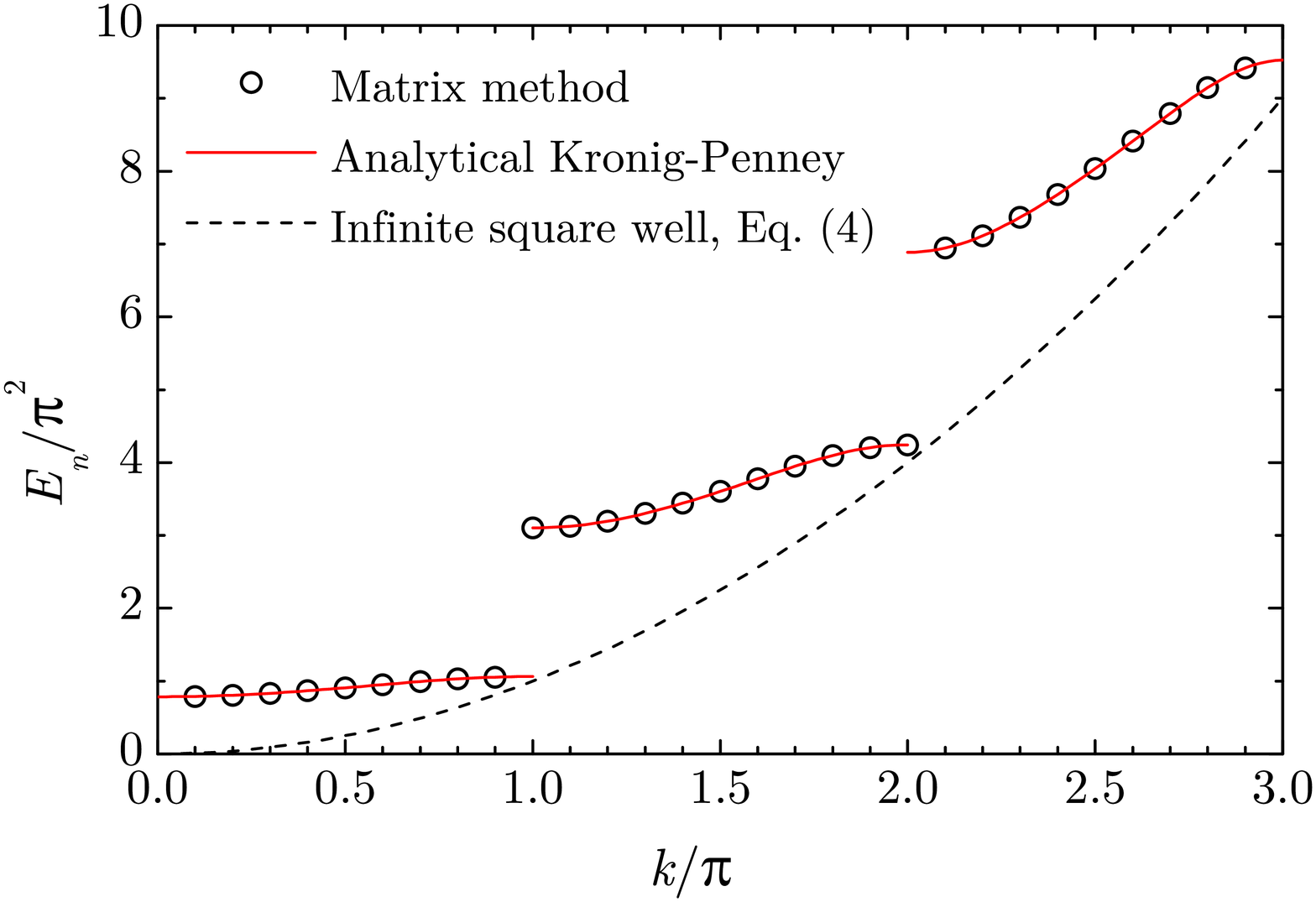}
\caption{Energies $E_n$ vs.\ wave numbers $k_n=n\pi/L$, for the one-dimensional Kronig-Penney potential with $b=1/6$ and $V_0=100$. Note the band structure. The matrix method values were obtained for $n_b=10$ and $N=100$.
}
\label{figBand10}
\end{figure}

Figure~\ref{figBand10} shows the energies $E_n$ for a periodic potential with barrier width $b=1/6$, $n_b=10$ barriers ($L=10$), and $V_0=100$, calculated using the matrix method with $N=100$. The continuous lines represent the energies calculated from the analytical Kronig-Penney solution.
This plot shows that the agreement between the matrix method and the analytical results, even for only ten barriers, is excellent. The difference between the two sets of data is of the order of 0.1\%.  We note that the error in the calculated energies depends on the ratio $n_b/N$, so that one should use larger $N$ values for larger $n_b$ values. Our results also agree with those obtained using a variant of the matrix formalism by Pavelich and Marsiglio,\cite{Pavelich15} who make explicit use of the Bloch theorem to build the wave functions for the periodic Kronig-Penney system.

The computational cost of such a matrix calculation is minimal by today's standards. For $n_b$ of the order of a few tens and $N$ a few hundreds, most of the computation time is devoted to the evaluation of the $N^2$ matrix elements $H_{nm}$, whereas the time required to find the eigenvalues and eigenvectors is negligible.\cite{SuplementaryMaterials} The matrix elements are calculated from the $2N$ quantities $F_{nm}(s+b/2)$ and $F_{nm}(s-b/2)$ defined above, so that the computation time is reduced by evaluating these in advance. On the other hand, Eq.~\eqref{HnnKP} indicates that the number of operations required to evaluate each matrix element scales as $n_b$. Therefore the time required by the matrix method to find the solutions scales as $n_b \,N^2$. Using \textsc{Mathematica} on a conventional personal computer, the calculation of the data of Fig.~\ref{figBand10} takes around three seconds.

\begin{figure}[h!]
\centering
 \includegraphics[width=8.5cm]{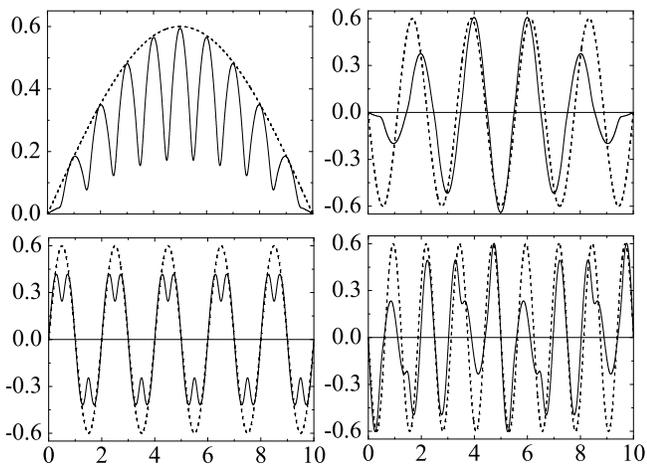}
\caption{Eigenfunctions $\psi_n(x)$ vs.~$x$ for (starting at top-left) $n=1$, $n=9$, $n=10$, and $n=16$, obtained by means of the matrix method for $b=1/6$, $V_0=100$, $n_b=10$, and $N=100$. The broken lines are the functions $0.6\sin(n\pi x/L)$. }
\label{figenvolventes}
\end{figure}

Notice that in Fig.~\ref{figBand10} we plotted the energy levels vs.\ a scaled version of the quantum number~$n$. Using solid-state physics nomenclature, this scaled quantum number is the wave number $k_n= n\pi/L$. It turns out that this wave number is actually related to the overall shape of the wave functions. In particular, the function $\sin(n\pi x/L)$ is a good approximation for the envelope of the wave function for the $n$th state, at least away from the square-well edges; this effect is illustrated in Fig.~\ref{figenvolventes} for selected values of~$n$.  These are the ``Bloch standing waves'' described by Johnston and Segal.\cite{Johnston1992}

\subsection{From energy levels to energy bands}

\begin{figure}[th!]
\centering
    \includegraphics[width=8.5cm]{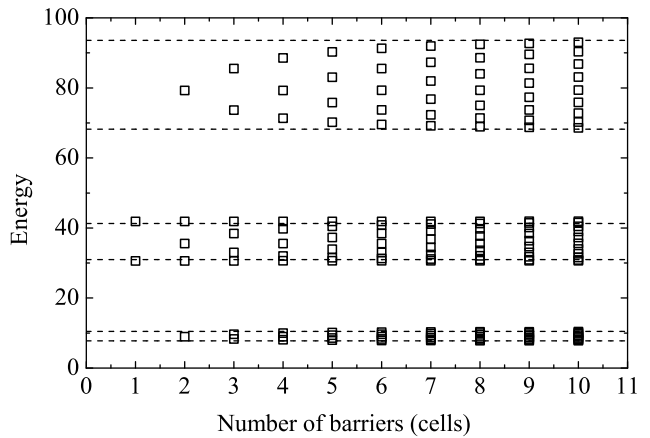}
\caption{Numerical matrix results (symbols) for the energy levels vs.~number of barriers (or cells) for configurations as in Fig.~\ref{fig1} with $b=1/6$, $V_0=100$, and $N=100$. The dashed lines correspond to the band limits of the original KP model.
}
\label{bandn}
\end{figure}

Here we show how energy bands appear as the number of internal barriers inside a square well is increased.
Our approach is similar to that employed by Cota \textit{et al.}\ for the KP model with Dirac delta barriers.\cite{Cota1988} Figure~\ref{bandn} shows the energy levels of the system formed by $n_b$ barriers of height $V_0=100$ and thickness $b=1/6$ placed, as shown in Fig.~\ref{fig1}, inside a infinite square well of width $L=n_b$.  The number of levels below $V_0$  increases with the number of unit cells, and they gather to yield intervals of allowed levels (the bands), separated by regions with no allowed energies (the gaps or forbidden energies).

\subsection{ Dimerized Kronig-Penney}
\label{secDimer}

\begin{figure}[th!]
	\centering
	\includegraphics[width=8.5cm,height=8cm]{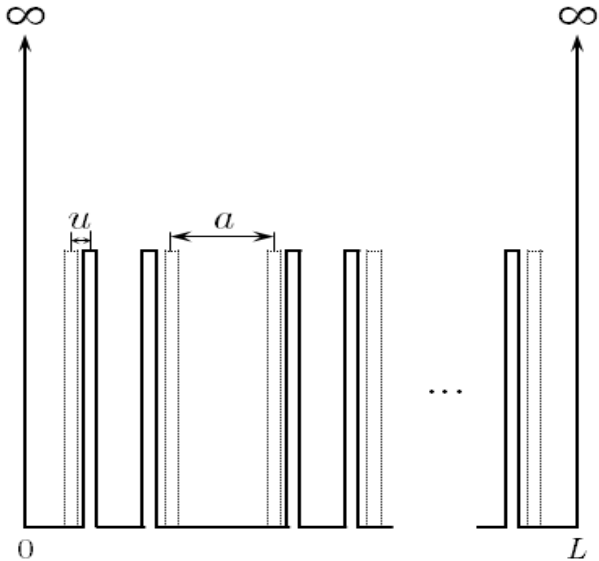}
	\caption{Potential for the dimerized Kronig-Penney model. Solid line: dimerized potential with parameter $u\neq 0$; dotted line: original ($u=0$) potential. }
	\label{figpotentialdimers}
\end{figure}

We consider now a \textit{dimerized} Kronig-Penney model in which the barriers are alternately shifted right or left by a distance~$u$ (see Fig.~\ref{figpotentialdimers}).  This model is equivalent to a KP model with a unit cell of length 2 (the dimer), containing two barriers placed at locations $(1/2)+u$ and $(3/2)-u$.

For the special case of Dirac delta barriers, the dimerized KP model has been studied by Go\~{n}i \textit{et al.}~\cite{Goni1986} The periodic potential is then $V(x)=2P\, \sum_r \delta(x-x_r)$, where $x_r=-1/2+r-(-1)^r u$ and $2P$ is a parameter that determines the strength of the barrier.  The analytical calculations of Ref.~\onlinecite{Goni1986} predict the appearance of energy gaps whose positions and widths depend on the dimerization parameter~$u$.

\begin{figure}[th!]
	\centering
	\includegraphics[width=8.5cm]{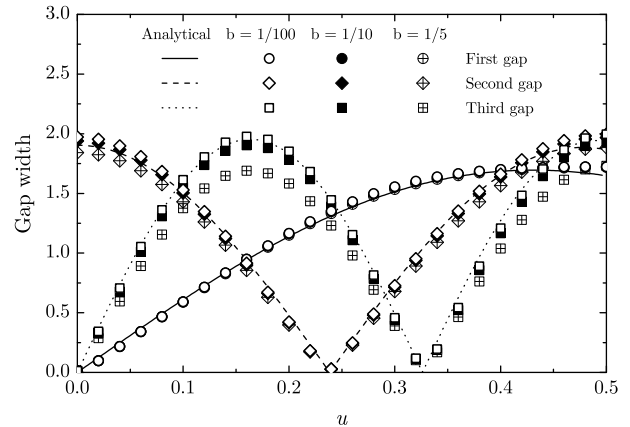}
	\caption{ The three first gap widths vs.~the dimerization parameter $u$ for the dimerized KP model. Open symbols: results from the matrix method for the first (circles), second (diamonds) and third (squares) gap, obtained with $b=1/100$, $V_0=100$ (open symbols), $b=1/10$, $V_0=10$ (filled symbols) and $b=1/5$, $V_0=5$ (crossed symbols). In all cases, $n_b=80$ and $N=200$. Solid lines: analytical results for the Dirac delta KP model.\cite{Goni1986}}
\label{figGaps}
\end{figure}

In order to compare these analytical results with those from the matrix method for barriers of width $b$ and height $V_0$, we must employ thin, high barriers ($b\to 0$ and $V_0\to \infty$ with the $b V_0=2P$ held fixed) that mimic Dirac delta barriers.\cite{Kronig1931,Wolfe1978} We have used $b=1/100$ and $V_0=100$ for the results displayed  as open symbols in Fig.~\ref{figGaps}. This figure shows good agreement between the theoretical gap widths and those calculated with the matrix approach.  In this respect, a minor technical point about our numerical estimate of the gaps between bands is in order. Let us assume that the extreme values of two contiguous bands are $E_n$ and $E_{n+1}$. Then our (improved) estimate of the size of the gap between the two bands is the difference between the extrapolation value at the middle point from the right, $E_{n+1}-(E_{n+2}-E_{n+1})/2$, and from the left, $E_{n}+(E_{n}-E_{n-1})/2$.

Figure~\ref{figGaps} also displays the numerical results obtained with $b=1/10$, $V_0=10$ (filled symbols), and with $b=1/5$, $V_0=5$ (crossed symbols). They show that the gaps shrink as the barriers thicken. It is also notable that for relatively thick barriers with $b=1/10$ the gaps are still close to those of delta-type barriers; however, the discrepancies are already important for $b=1/5$. Note finally that the differences with the behavior for delta barriers increase with the order of the gaps.

\subsection{Surface electronic states in the Kronig-Penney model}
\label{secSurfaceState}

\begin{figure}[th!]
	\centering
	\includegraphics[width=8.5cm,height=7.5cm]{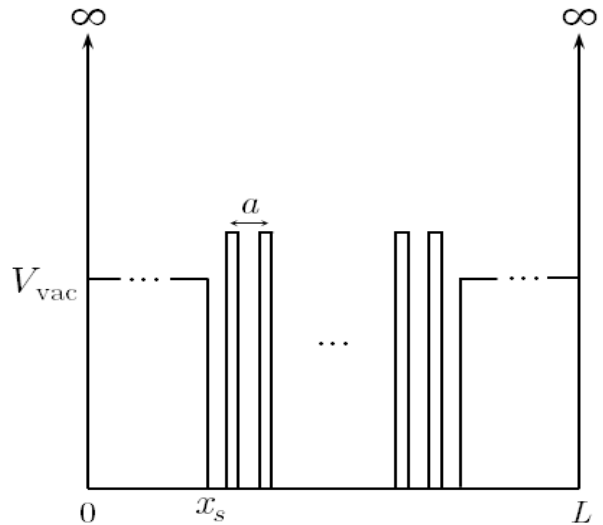}
	\caption{
The potential for a finite Kronig-Penney solid with two surfaces of height  $V_\text{vac}$  at~$x_s$ and $L-x_s$.
}
\label{figpotentialsurface}
\end{figure}

Next we modify the original Kronig-Penney model to add a ``surface'' that can cause the appearance of  so-called (Tamm) surface states.\cite{Tamm1932,Davison1996,Steslicka1974,Steslicka1966}  This surface is represented by a potential $V_\text{vac}$ that the electron has to surmount to escape from the crystal to the vacuum (as shown in Fig.~\ref{figpotentialsurface}).  Since surfaces constitute local breakdowns of the translational symmetries of ideal solids, the wave number $k$ appearing in the $e^{ikx}$ factor of the Bloch waves, which is always real for ideal crystals, may, in some cases, be complex,\cite{Davison1996,Steslicka1966} leading to wave functions localized near the surface with energies inside the forbidden energy gaps of the ideal infinite crystal.

A model amenable to a relatively simple analytical description is the (semi-infinite) KP model with an \textit{infinite} number of equidistant Dirac delta barriers placed to the right of a surface (vacuum) represented by a constant potential $V_\text{vac}$. That is,
\begin{equation}
\label{11}
V(x)=V_\text{vac}\, \theta(x_s-x)+2P\, \sum_{r=1}^\infty \delta(x-x_r),
\end{equation}
where $\theta(x)$ is the Heaviside step function, $2P$ is the strength of the delta barriers, $x_s$ is the position of the surface, and $x_r=x_s-1/2+r$ are the positions of the barriers.
For the fully infinite KP ideal crystal (without surfaces) the dispersion relation is well known:\cite{Griffiths2004,Kronig1931,Wolfe1978,Davison1996}
\begin{equation}
\label{drKP}
\cos k=\cos\xi + P\xi^{-1} \sin\xi,
\end{equation}
where  $\xi^2=E$. Here $k$ must be real for the Bloch wave function to remain finite. Shortly after the Kronig-Penney work,\cite{Kronig1931} Tamm\cite{Tamm1932} realized that this requirement no longer holds for a semi-infinite crystal (defined, for example, as in Eq.~\eqref{11}), and that solutions with complex wave numbers of the form $k=i\beta+m \pi$, with $\beta>0$ real and $m=0,1,\ldots\,$, may exist.
In these cases, the energies $E=\xi^2$ of the surface states for a KP lattice with delta barriers of strength $2P$ and a surface of height $V_\text{vac}=\xi_0^2$ are\cite{Davison1996,Steslicka1966,Steslicka1973}
\begin{equation}
\label{12}
\xi \cot\xi=\frac{\xi_0^2}{2P} - \sqrt{\xi_0^2-\xi^2},
\end{equation}
provided that $\beta>0$. These energies are inside the forbidden energy gaps.
In order to show the space localization of the surface states it is convenient to define the relative probability density\cite{Thakkar1978}
$R(x)= \left|\psi^{{s}}(x)\right|^2/\left|\psi^{{s}}(0)\right|^2$, where $\psi^{{s}}(x)$ is a surface state wave function.

It turns out that no surface state can exist unless its energy satisfies the so-called \emph {Tamm existence condition},\cite{Steslicka1973}
$\xi_0^2 <  \xi^2+P^2$,
which using Eq.~\eqref{12},
can be restated in an equivalent way: surface states can only exist for vacuum potentials $V_\text{vac}$ smaller than the limit defined by
\begin{equation}
\label{Vvaclimit}
(\xi_0^2-P^2)^{1/2} \; \cot(\xi_0^2-P^2)^{1/2}=\frac{\xi_0^2}{2P}-P.
\end{equation}
For $V_\text{vac}\to\infty$ the Tamm existence condition is never satisfied, no forbidden-energy-gap state appears, and only standard KP energy bands coming from Eq.~\eqref{drKP} remain.

A similar analytical treatment for the case of \textit{two} surfaces, to the left and right of a finite train of equidistant Dirac delta barriers, is possible but more involved,\cite{Davison1996,Steslicka1974} and will not be considered here. However, it turns out that the preceding analysis for the infinite system provides an accurate description for even relatively small finite systems,\cite{Davison1996} which we will use in our matrix approach.  Physically, the effects of a surface at the right on the phenomena occurring at the left are negligible if the two surfaces are far apart.

The study of surface states in the finite KP model (i.e., with two surfaces)
by means of the matrix method is straightforward: in the box of length $L$ we place $n_b=L-3$ barriers of width $b$ and height $V_0$ at positions $x_r=r$, with $r=2,3,\ldots,n_b+1$, and then attach a barrier of height $V_\text{vac}$ and width $1+b/2$ to each infinite wall, as shown in Fig.~\ref{figpotentialsurface}. The position of the left surface is then $x_s=1+b/2$. This way the widths $1-b$ of all the $n_b+1$ valleys are the same, just to mimic the model with Dirac delta barriers. The matrix elements $H_{nm}$ are then readily obtained (\emph{cf}.~Eq \eqref{HnnKP}):
\begin{align}
\label{HnnKPS}
H_{nm}&= E_n^{(0)} \delta_{nm}+ V_0\, \sum_{r=2}^{n_b+1} h_{nm}\left(x_r,b\right)\nonumber\\
 &+V_\text{vac} \left[ h_{nm}\left(x_L,1+b/2\right)
 + h_{nm}\left(x_{R},1+b/2\right)\right],
\end{align}
with $x_L=x_s/2$ and $x_R=L-x_L$.

\begin{figure}[th!]
\centering
    \includegraphics[width=8.5cm]{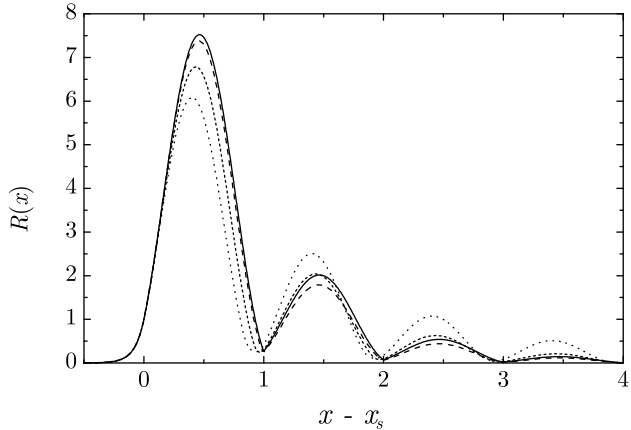}
\caption{ Results of the matrix method for the relative probability density $R(x)=\left|\psi^{{s}}_1(x)\right|^2/\left|\psi^{{s}}_1(0)\right|^2 $ of the first surface state with $P=10$, $V_\text{vac}=50$, $N=400$, $n_b=10$, and $b=1/6$ (dotted line), $b=1/12$ (short-dashed line), and $b=1/96$ (dashed line). The solid line is the theoretical result for Dirac delta barriers.\cite{Davison1996} }
\label{figSurfaceState}
\end{figure}

Figure~\ref{figSurfaceState} plots the relative probability density $R(x)$ corresponding to the first surface state (i.e., $\left|\psi^{{s}}_1(x)\right|^2/\left|\psi^{{s}}_1(0)\right|^2 $) of the semi-infinite Dirac delta KP model with $P= 10$ and $V_\text{vac}=50$.\cite{Davison1996}
The analytical model through  Eq.\eqref{12} yields only two surface states, with energies $E^{{s}}_1=6.65$ and $E^{{s}}_2=26.44$, for this case.
Figure~\ref{figSurfaceState} also shows the corresponding $R(x)$ as obtained by the matrix method with $N=400$ for $n_b=10$ barriers of thickness $b=1/6$, $b=1/12$, and $b=1/96$, with $V_0 = 2P/b$.
For finite crystals, each surface energy is actually split into a pair of values $\{E_n^{s,a},E_n^{s,b}\}$, which coalesce into the infinite crystal value $E_n^{s}$ when $n_b\to \infty$.\cite{Davison1996}  Table~I lists both energies in each case for $n_b=10$, and for comparison, also lists the (coalesced) energy for $n_b=20$.    The agreement of the numerical matrix results with the theoretical ones corresponding to the infinite KP model with Dirac delta barriers increases as we reduce the thickness of the barriers, as expected.
\begin{table}[h!]
\centering
\caption{Surface state energies for the Kronig-Penney solid shown in Fig.~7, with $V_\text{vac}=50$ and $N=400$.
For the sake of simplicity, we only write down a single value for $n_b=20$ because the values of $E_n^{s,a}$ and $E_n^{s,b}$ rounded up to the hundredth are equals. }
\begin{ruledtabular}
\begin{tabular}{l c c c c }
$b$ & $E_1^s$ ($n_b=10$) & $E_1^s$ ($n_b=20$) & $E_2^s$ ($n_b=10$) & $E_2^s$ ($n_b=20$) \\
\hline
1/6  & 8.22,\,8.23 & 8.23 & 31.91,\,31.92 & 31.92 \\
1/12 & 7.37,\,7.37 & 7.37 & 28.99,\,29.00 & 29.00 \\
1/96 & 6.77,\,6.78 & 6.83 & 26.86,\,26.86 & 27.02 \\
\end{tabular}
\end{ruledtabular}
\label{SurfaceEnergiesTable}
\end{table}

According to Eq.~\eqref{Vvaclimit}, the maximum value of $V_\text{vac}$ that supports a surface state is  $V_{\text{vac}}\approx 107$ for the KP model with Dirac delta barriers with $P=10$. For larger values of $V_{\text{vac}}$ the surface state leaves the forbidden energy gap, entering an allowed energy band, and the wave function loses its damped behavior. For $b = 1/96$, $P=10$, $n_b=10$, and $N=400$,
the limiting $V_{\text{vac}}$ value that one finds numerically\cite{SuplementaryMaterials} is around  $110$, in good agreement with the value $V_{\text{vac}}\approx 107$ of the Dirac KP model.
  Thus, the value  $V_{\text{vac}} = 50$ that we have chosen ensures the existence of surface states.
 Besides, this is a sensible value for the surface potential felt by an electron: for a lattice parameter $a=4$ {\AA},  $V_{\text{vac}} = 50$ is equivalent to roughly $12$ eV, which is a reasonable vacuum potential value.

\subsection{Kronig-Penney model with an external field}
\label{secKPextfield}

\begin{figure}[th!]
	\centering
	\includegraphics[width=8.5cm,height=8cm]{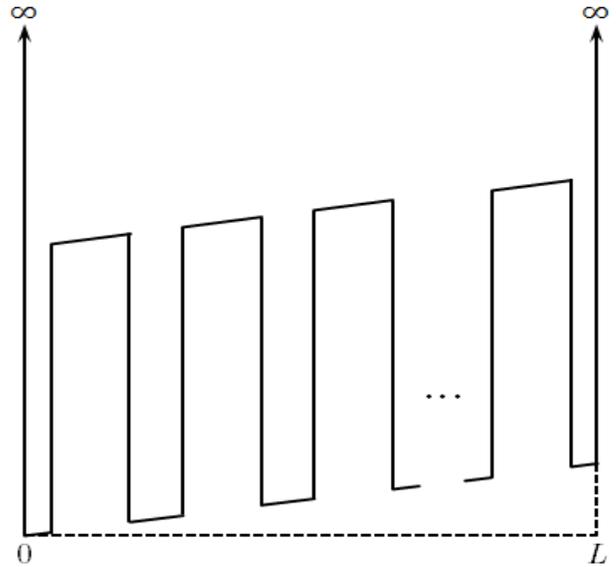}
	\caption{Potential for a Kronig-Penney solid with a constant electric field~$\epsilon$.}
	\label{figpotentialfield}
\end{figure}

Finally, we consider the problem of a finite Kronig-Penney solid in the presence of a uniform electric field $-\epsilon$, that is,
$V_\text{e}(x)=\epsilon x+V_\text{inf}(x)+V_{KP}(x)$,
where $V_{KP}(x)$ is given by Eq.~\eqref{Vaux}. The shape of
$V_\text{e}(x)$ is shown in Fig.~\ref{figpotentialfield}.
 The matrix elements $H_{nm}$ are just those of the KP model, $H^{KP}_{nm}$, plus the contribution $H^{e}_{nm}$ from the external field:
\begin{eqnarray}
H^{e}_{nm} &=&
\frac{2\epsilon}{L} \int_0^L\,\sin\left(\frac{n\pi x}{L}\right)\,x\, \sin\left(\frac{m\pi x}{L}\right) \, dx \nonumber\\
&=&\label{Henm}
\begin{cases}
\dfrac{\epsilon L}{2}, & m=n,\\
0, & m+n=\text{even},\\
-\dfrac{ 8 m n\, \epsilon \,L}{\pi ^2 \left(m^2-n^2\right)^2},&m+ n=\text{odd}.
\end{cases}
\end{eqnarray}

\begin{figure}[th!]
\centering
    \includegraphics[width=8.5cm]{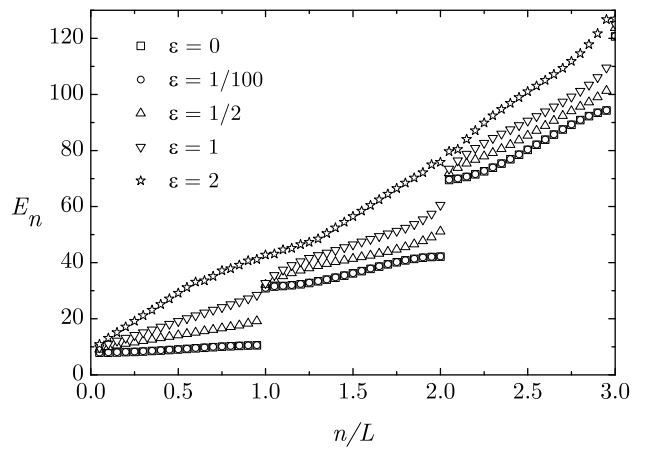}
\caption{ Band structure for several values of the field strength $\epsilon$ for a KP solid with $b=1/6$, $V_0=100$, $n_b=20$, and $N=100$. }
\label{figBandasCampo}
\end{figure}

\begin{figure}[th!]
\centering
    \includegraphics[width=8.5cm]{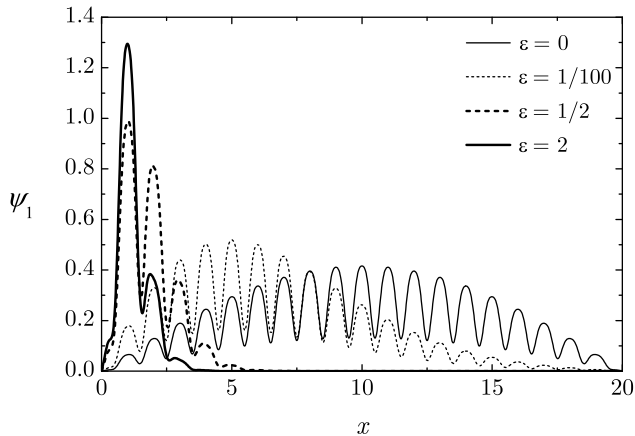}
\caption{ Wave functions
 corresponding to the lowest allowed energy for the KP solid with an electric field for $b=1/6$, $V_0=100$, $n_b=20$, $N=100$, and several field strengths.
}
\label{figOndasCampo}
\end{figure}

The numerically calculated energy bands for different values of the field strength are shown in Fig.~\ref{figBandasCampo};  wave functions corresponding to the lowest allowed energy for the same fields appear in Fig.~\ref{figOndasCampo}. In all cases we have used $b=1/6$, $V_0=100$, $n_b=20$, and  $N=100$.

The first obvious effect of the field consists of the reduction of the forbidden band widths with increasing the field strength;
eventually, for fields intense enough, the forbidden bands disappear. The low-energy wave functions shift towards the region of lower potential, with a larger shift for larger field intensities.
It is remarkable how little the energy bands change with the external field in comparison to what happens for the wave functions; compare, for example, the energy bands and wave functions $\psi_1(x)$ for $\epsilon=0$ and $\epsilon=1/100$.

 A complementary discussion of the behavior of the bands and wave functions of the Kronig-Penney model with an external field can be found in Ref.~\onlinecite{Johnston1992}.

\subsection{Additional problems}

Many other problems, similar to those discussed above, could be considered.

For instance, the effect of point defects (and complex associations) in the band structure of crystals can be studied by modifying the widths and heights of some barriers in the Kronig-Penney potential; resonant levels within bands or localized electronic states within the forbidden energy gaps should appear in this case (see Sec.~III.D-E in Ref.~\onlinecite{Johnston1992}).

As simple models of amorphous materials, one could study disordered KP models where the heights and/or widths and/or separations of the barriers are random.\cite{Saperstein1983,Johnston1992}

In Sec.~\ref{secSurfaceState}, numerical results for the surface states of finite crystals were compared with the corresponding analytical results for the infinite crystal, but a further comparison could be carried out between numerical and analytical results for the finite crystal (see, for example, Sec.~3.3 of Ref.~\onlinecite{Davison1996} for the theoretical discussion of this case), for a crystal with a distorted surface, \cite{Steslicka1966,Neuberger1975} or with an external field.\cite{Thakkar1978}

Energy bands for non-rectangular periodic potentials (such as the Mathieu sinusoidal potential) can also be studied.\cite{Pavelich15}

\begin{figure}[h!]
\centering
\includegraphics[width=8.5cm]{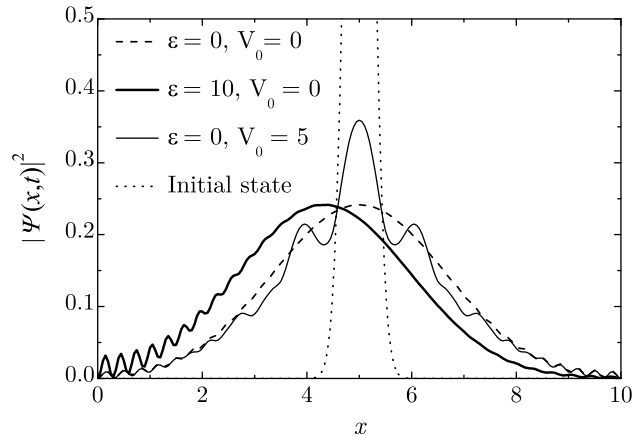}
\caption{ Density of probability $\left|\Psi(x,t)\right|^2$ at time $t=0.26$ for several values of $\epsilon$ and $V_0$. The initial state \eqref{wavepacket} was calculated with $\sigma^2=0.05$ and $x_0=L/2$. In all cases $b=1/6$, $n_b=10$ and $N=100$. We use a time unit where $\hbar=1$. }
\label{figpsixt}
\end{figure}

Finally, the matrix approach provides a straightforward way to compute the time evolution of some quantum states. Assuming that $\Psi(x,0)\approx \sum_{n=1}^N a_n \psi_n(x) $ is a fair approximation to the initial quantum state, then we know that\cite{Belloni2008}
\begin{equation}
\label{Psixt}
\Psi(x,t)\approx
\sum_{n=1}^N a_n\, e^{-iE_nt/\hbar}\, \psi_n(x).
\end{equation}
We have seen in previous sections that the matrix method provides good approximations for the eigenvalues $E_n$ and eigenfunctions $\psi_n(x)=\sum_{m=1}^N c_m^{(n)} \varphi_m(x)$ appearing in this expression. Then, in order to evaluate Eq.~\eqref{Psixt}, all that one needs is to compute the coefficients $a_n$:
\begin{equation}
\label{acoef}
a_n=\langle\psi_n|\Psi(x,0)\rangle
\approx \sum_{m=1}^N c_m^{(n)*}\,\langle\varphi_m|\Psi(x,0)\rangle,
\end{equation}
where
\begin{equation}
\label{acoefExpli}
\langle\varphi_m|\Psi(x,0)\rangle= \sqrt{\frac{2}{L}}\int_0^L \sin\left(\frac{m \pi x}{L}\right) \Psi(x,0) \,dx.
\end{equation}
As an example, in Fig.~\ref{figpsixt} we show the probability density $\left|\Psi(x,t)\right|^2$ obtained from Eqs.~\eqref{Psixt} through \eqref{acoefExpli}, for an initial Gaussian wave packet
\begin{equation}
\label{wavepacket}
\Psi(x,0)= \frac{1}{\left(\pi \sigma^2\right)^{1/4}}\exp\left[-\frac{(x - x_0)^2}{2 \sigma^2}\right],
\end{equation}
for three different cases of the system of Sec.~\ref{secKPextfield}, namely a case with no barriers and no external field, a case with barriers and no external field, and a case with external field but no barriers.
Another interesting task would be to study how the position of the wave packet changes according to the value of the external field and/or size of the barriers. For example, for the case with $\epsilon=10$ and $V_0=0$, it is a worthwhile exercise to check that the movement of the location of the peak of the wave packet, $x_{\text{max}}$, shown in Fig.~\ref{figpsixt} from $t=0$ to $t=0.26$, obeys Newton's second law, $x_{\text{max}}=x_0- \epsilon t^2/2\mu$.

\section{Summary}

 In this paper we have shown that the numerical matrix procedure described by Marsiglio\cite{Marsiglio2009} can be easily applied to periodic potentials that model a variety of interesting phenomena in solid state physics. The matrix method is especially appropriate for these systems because it conveniently yields, with high accuracy, a whole list of eigenvalues (e.g., the energy bands) and the associated eigenfunctions. It is also simple to program and use, and quite efficient computationally.  We have employed this procedure to show that some characteristics of fully periodic systems, such as the existence of energy bands and forbidden intervals, appear already when one considers a relatively small number of unit cells. We have illustrated the capability of the method by applying it to the Kronig-Penney (KP) potential and some related systems, such as a dimerized KP solid, a KP solid with surfaces, and a KP solid with an external electric field. Finally, we have shown that the method can be readily employed to obtain the time evolution of quantum states.

\begin{acknowledgments}

This work was partially funded by the Ministerio de Ciencia y Tecnolog\'ia (Spain) through Grant No.\  FIS2013-42840-P (partially financed by FEDER funds), by the Ministerio de Econom\'ia e Innovaci\'on (Spain) under Grant No. MAT2012-38205-C02-02, and by the Junta de Extremadura through Grant No.\  GR10158.
We thank the editorial staff of the American Journal of Physics for help in the improvement of this manuscript.

\end{acknowledgments}


\begin{thebibliography}{99}

\bibitem{Kinderman1990} J. V. Kinderman, ``A computing laboratory for introductory quantum mechanics,'' Am. J. Phys. \textbf{58}, 568--573 (1990).

\bibitem{French1979} A. P. French and E. F. Taylor, \textit{An Introduction to Quantum Physics} (Norton, New York, 1978).

\bibitem{Chow1972} P. C. Chow, ``Computer Solutions to the Schr{\"o}dinger Equation,''
 Am. J. Phys. \textbf{40}, 730--734 (1972).

\bibitem{Johnston1992}  I. D. Johnston and D. Segal, ``Electrons in a crystal lattice: A simple computer model,'' Am. J. Phys. \textbf{60}, 600--607 (1992).

\bibitem{Marsiglio2009} F. Marsiglio, ``The harmonic oscillator in quantum mechanics: A third way,'' Am. J. Phys. \textbf{77}, 253--258 (2009).

\bibitem{Pavelich15}  R. L. Pavelich and F. Marsiglio, ``The Kronig-Penney model extended to arbitrary potentials via numerical matrix mechanics,''
    Am. J. Phys. \textbf{83}, 773--781 (2015).

\bibitem{Jugdutt2013} B. A. Jugdutt and F. Marsiglio, ``Solving for three-dimensional central potentials using numerical matrix methods,'' Am. J. Phys. \textbf{81}, 343--350 (2013).

\bibitem{Griffiths2004}
	D. J. Griffiths, \textit{Introduction to Quantum Mechanics}, 2nd ed.\ (Pearson Prentice Hall, Upper Saddle River, NJ, 2004).

\bibitem{SuplementaryMaterials} See supplementary material at
\footnotesize{\url{<http://www.eweb.unex.es/eweb/fisteor/santos/KP.zip>}} where
{\footnotesize MATHEMATICA} codes demonstrating how our calculations are carried out are available.

\bibitem{Kronig1931}
	R. de L. Kronig and W. G. Penney, ``Quantum Mechanics of Electrons in Crystal Lattices,'' Proc. R. Soc. London A \textbf{130}, 499--513 (1931).

\bibitem{Titus1973}
 W. J. Titus, ``Solutions of Kronig-Penney Models by the T-Matrix Method,'' Am. J. Phys. \textbf{41}, 512--516 (1973); G. C. Wetsel, Jr., ``Calculation of the energy-band structure of the Kronig-Penney model using the nearly-free and tightly-bound-electron approximations,'' Am. J. Phys. \textbf{46}, 714--720 (1978).

\bibitem{Singh1982}
 S. Singh, ``Kronig-Penney model in reciprocal lattice space,'' Am. J. Phys. \textbf{51}, 179 (1983); 	 F. Szmulowicz, ``Kronig-Penney model: a new solution,'' Eur. J. Phys. \textbf{18}, 392--397 (1997).


\bibitem{Lippmann1997} H. Lippmann, ``Remarks about the manipulation of the Kronig-Penney model for the introduction into the energy band theory of crystals,'' Am. J. Phys. \textbf{65}, 89--92 (1997).

\bibitem{Szmulowicz1997} F. Szmulowicz, ``New eigenvalue equation for the Kronig-Penney problem,'' Am. J. Phys. \textbf{65}, 1009--1014 (1997).

\bibitem{Szmulowicz2008} F. Szmulowicz, ``New Kronig--Penney equation emphasizing the band edge conditions,'' Eur. J. Phys. \textbf{29}, 507--515 (2008).

\bibitem{Cota1988} E. Cota, J. Flores, and G. Monsivais, ``A simple way to understand the origin of the electron band structure,'' Am. J. Phys. \textbf{56}, 366--372 (1988).

\bibitem{Goni1986} A. R. Go\~{n}i, A. G. Rojo, and E. N. Mart\'{\i}nez, ``A dimerized Kronig-Penney model,'' Am. J. Phys. \textbf{54}, 1018--1021 (1986).

\bibitem{Wolfe1978}  J. C. Wolfe, ``Summary of the Kronig-Penney electron,'' Am. J. Phys. \textbf{46}, 1012--1014 (1978).

 \bibitem{Tamm1932} I. E. Tamm, ``\"{U}ber eine m\"{o}gliche Art der Elektronenbindung an Kristalloberf\"{a}chen,'' Phys. Z. Sowjetunion  \textbf{1},  733--746 (1932).

 \bibitem{Davison1996}
 S. G. Davison and M. St\k{e}\'{s}licka, \textit{Basic Theory of Surface States} (Oxford University Press, Oxford, 1996).


\bibitem{Steslicka1974} M. St\k{e}\'{s}licka, ``Kronig-Penney model for surface states,'' Prog. Surf. Sci. \textbf{5}, 157--259 (1974).


\bibitem{Steslicka1966}  M. St\c{e}\'{s}licka and  K. F. Wojciechowski, ``Surface states of a deformed one-dimensional crystal,'' Physica \textbf{32}, 1274--1282 (1966).

\bibitem{Neuberger1975}  J. Neuberger and C. R. Fischer, ``Tamm states at a distorted surface,'' Physica B \textbf{79}, 350--358 (1975).

\bibitem{Steslicka1973}  M. St\c{e}\'{s}licka, ``Note on the existence conditions of surface states,'' Phys. Lett. A \textbf{44}, 513--514 (1973).

\bibitem{Thakkar1978} A. J. Thakkar and  M. St\c{e}\'{s}licka, ``Model studies of the Tamm-like and field-sustained surface states of germanium,'' Surf. Sci. \textbf{74}, 168--180 (1978).

\bibitem{Saperstein1983} A. M. Saperstein, ``Energy gaps in one-dimensional amorphous materials: A disordered Kronig-Penney model,'' Am. J. Phys. \textbf{51}, 1127--1130 (1983).

\bibitem{Belloni2008} M. Belloni and W. Christian, ``Time development in quantum mechanics using a reduced Hilbert space approach,'' Am. J. Phys. \textbf{76}, 385--392 (2008). These authors have built a suite of open-source programs that employ Eq.~\eqref{Psixt} to calculate and visualize the time evolution of arbitrary bound states.

\end{thebibliography}
\end{document}